\documentclass[aps,prc]{revtex4}

\usepackage{graphics}
\usepackage{epsfig}
\usepackage{amsmath}

\newcommand{\ba}{\begin{eqnarray}}
\newcommand{\ea}{\end{eqnarray}}

\begin{document}

\title{Space- and time-like electromagnetic form factors of the nucleon}
\author{R. Bijker}
\affiliation{Instituto de Ciencias Nucleares, 
Universidad Nacional Aut\'onoma de M\'exico, 
Apartado Postal 70-543, 04510 M\'{e}xico, D.F., M\'{e}xico
\footnote{Invited talk at 'XXVIII Nuclear Physics Symposium', 
Cocoyoc, M\'exico, January 4-7, 2005}}
\date{January, 2005}

\begin{abstract}
Recent experimental data on space- and time-like form factors of the 
nucleon are analyzed in terms of a two-component model with a quark-like 
intrinsic structure and a meson cloud. A good overall agreement is found 
for all electromagnetic form factors with the exception of the neutron 
magnetic form factor in the time-like region. 

\

\end{abstract}

\maketitle

\section{Introduction}

The structure of the nucleon is of fundamental importance in 
nuclear and particle physics. The electromagnetic form factors 
are key ingredients to the understanding of the internal structure 
of composite particles like the nucleon since they contain the 
information about the distributions of charge and magnetization.
The first evidence that the nucleon is not a point particle but has 
an internal structure came from the measurement of the anomalous 
magnetic moment of the proton in the 1930's \cite{stern}, which 
was determined to be 2.5 times as large as one would expect for 
a spin $1/2$ Dirac particle (the actual value is 2.793). The finite 
size of the proton was measured in the 1950's in electron scattering 
experiments at SLAC to be $\sim 0.8$ fm \cite{hofstadter} (compared to the 
current value of 0.895 fm). The first evidence for point-like constituents 
(quarks) inside the proton was found in deep-inelastic-scattering 
experiments in the late 1960's by the MIT-SLAC collaboration \cite{dis}, 
which eventually together with many other developments would lead to the 
formulation of QCD in the 1970's as the theory of strongly interacting 
particles. 

The complex structure of the proton manifested itself once again in 
recent polarization transfer experiments \cite{jones,gayou} which showed 
that the ratio of electric and magnetic form factors of the proton 
exhibits a dramatically different behavior as a function of the momentum 
transfer as compared to the generally accepted picture of form factor 
scaling obtained from the Rosenbluth separation method \cite{walker,andivahis}. 
The discrepancy between the experimental results has been the subject of 
many theoretical investigations which have focussed on radiative corrections  
due to two-photon exchange processes \cite{twophoton,tomasi}.  

The new experimental data for the proton form factor ratio are in excellent 
agreement with a phenomenological model of the nucleon put forward in 1973 
\cite{IJL} wherein the external photon couples both to an intrinsic structure 
and to a meson cloud through the intermediate vector mesons ($\rho$, $\omega$, 
and $\varphi$). The linear drop in the proton form factor ratio was also 
predicted in a chiral soliton model \cite{soliton} before the polarization 
transfer data from Jefferson Lab became available in 2000 \cite{jones,gayou}. 
On the contrary, the new experimental data for the neutron \cite{madey} 
are in agreement with the vector meson dominance (VMD) model of \cite{IJL} 
for small values of $Q^{2}$, but not so for higher values of $Q^2$. 

The aim of this contribution is to present an simultaneous study of all 
electromagnetic form factors of the nucleon not only in the space-like region 
where they can be measured in electron scattering experiments, but also in the 
time-like region where they can be studied through the creation or 
annihilation of a nucleon-antinucleon pair. The analysis is carried out 
in a modified version of \cite{IJL}: a two-component model consisting of an 
intrinsic (three-quark) structure and a meson cloud whose effects are taken 
into account via vector meson dominance (VMD) couplings \cite{BI}. 

This manuscript is organized as follows: in Section~2 the experimental situation 
for the form factor ratio of the proton is reviewed briefly. The main 
ingredients of the two-component model of \cite{BI} and its application 
to the space-like form factors are discussed in Section~3 and extended 
to the time-like region in Section~4. The last section contains the summary 
and conclusions.

\section{Electromagnetic form factors}

Electromagnetic form factors provide important information on the 
structure of the nucleon. Relativistic invariance determines the form 
of the nucleon current for one-photon exchange to be
\ba
J_{\mu} &=& F_1(Q^2) \, \gamma^{\mu} 
+ \frac{1}{2M} F_2(Q^2) \, i \sigma^{\mu\nu} q_{\nu} ~,
\ea
where $M$ denotes the nucleon mass, $q=(\omega,\vec{q})$ is 
the four-momentum of the virtual photon and $Q^2=-q^2$.  
$F_{1}$ denotes the Dirac form factor, and $F_{2}$ represents 
the helicity flip Pauli form factor which is proportional 
to the anomalous magnetic moment. The Sachs form factors, $G_{E}$ and 
$G_{M}$, can be obtained from $F_{1}$ and $F_{2}$ by the relations 
\ba
G_E(Q^2) &=& F_1(Q^2) - \tau F_2(Q^2) ~, 
\nonumber\\
G_M(Q^2) &=& F_1(Q^2) + F_2(Q^2) ~, 
\ea
with $\tau=Q^2/4 M^2$. $G_{E}$ and $G_{M}$ are the form factors describing the 
distribution of electric charge and magnetization as a function of the $Q^2$ 
satisfying 
\ba
G_{E_{p/n}}(0) &=& e_{p/n} ~, 
\nonumber\\
G_{M_{p/n}}(0) &=& \mu_{p/n} ~,
\ea
where $e_{p/n}$ and $\mu_{p/n}$ denote the electric charge and magnetic moment 
of the proton/neutron.  

\subsection{Rosenbluth separation method}

The differential cross section for elastic electron-nucleon scattering is given 
by the Rosenbluth formula \cite{rosenbluth}
\ba
\frac{d \sigma}{d \Omega} &=& \left( \frac{d \sigma}{d \Omega} \right)_{\rm Mott} 
\left[ \frac{G_E^2 + \tau G_M^2}{1 + \tau} + 2\tau G_M^2 \tan^2 \frac{\theta}{2} \right] 
\nonumber\\
&=& \left( \frac{d \sigma}{d \Omega} \right)_{\rm Mott} \frac{\tau}{\epsilon(1+\tau)}  
\left[ G_M^2 + \frac{\epsilon}{\tau} G_E^2 \right] ~,
\label{dcs}
\ea
where $\epsilon=1/[1+2(1+\tau)\tan^2 (\theta/2)]$ is the linear polarization of the 
virtual photon and $\theta$ the scattering angle.  
For a point-like proton ($G_E=G_M=1$) this expression reduces to the 
differential cross section for electron scattering off a spin-$1/2$ Dirac particle. 
By measuring the differential cross section at a fixed value of $Q^2$ as a function 
of $\epsilon$ (or equivalently, as a function of the scattering angle $\theta$), 
the electric and magnetic form factors can be disentangled according to the 
Rosenbluth separation method.
However, for large values of $Q^2$ the measured cross section is dominated by the 
magnetic form factor, which makes the determination of the electric form factor   
difficult due to an increasing systematic uncertainty with $Q^2$. 
The electric and magnetic form factors obtained with the Rosenbluth method 
can be described to a reasonable approximation (up to the 10-20 $\%$ level) by 
the so-called dipole fit 
\cite{bosted}
\ba
G_{E_p}(Q^2) \;\approx\; \frac{G_{M_p}(Q^2)}{\mu_p} \;\approx\; 
\frac{G_{M_n}(Q^2)}{\mu_n} \;\approx\; 
G_D(Q^2) \;=\; \frac{1}{(1+Q^2/0.71)^2} ~. 
\label{dipole}
\ea
The electric form factor of the neutron is relatively small and was  
parametrized in 1971 by the the Galster formula 
\ba
G_{E_n}(Q^2) \;=\; -\mu_n \frac{a \tau}{1+b\tau} G_D(Q^2) ~,
\ea
with $a=1$ and $b=5.6$ \cite{galster}. 
In a more recent fit, the coefficients $a$ and $b$ were determined to be 
$a=0.888 \pm 0.023$ and $b=3.21 \pm 0.33$ \cite{madey}.

\subsection{Polarization transfer method}

In recent years, experiments with polarized electron beams have become 
feasible  at MIT-Bates, MAMI and Jefferson Lab. In polarization transfer 
experiments 
\ba
\vec{e} + p \rightarrow e + \vec{p}  ~,
\ea
the polarization of the recoiling proton is measured from which 
the ratio of the electric and magnetic form factors can be determined 
directly as 
\ba
\frac{G_E}{G_M} &=& - \frac{P_t}{P_l} \frac{E + E'}{2M_p} 
\tan \frac{\theta}{2} ~, 
\label{ratio}
\ea
where $P_t$ and $P_l$ denote the polarization of the proton 
perpendicular and parallel to its momentum in the scattering plane.  
$E$ and $E'$ denote the initial and final electron energy.   
The simultaneous measurement of the two polarization components 
greatly reduces the systematic uncertainties. 
The results of the polarization transfer method showed a big surprise. 
The ratio of electric and magnetic form factors of the proton dropped 
almost linearly with $Q^2$ \cite{jones,gayou}
\ba
R_p \;=\; \frac{\mu_p G_{E_p}(Q^2)}{G_{M_p}(Q^2)} 
\;\approx\; 1 - 0.13 (Q^2 - 0.29) ~,
\label{rp2} 
\ea
in clear disagreement with the Rosenbluth results of Eq.~(\ref{dipole}) 
which correspond to a constant value of $R_p$  
\ba
R_p \;\approx\; 1 ~.
\label{rp1}
\ea
The vector meson dominance (VMD) model proposed in 1973 by 
Iachello, Jackson and Lande \cite{IJL} is in excellent agreement with 
the new polarization transfer data, although {\it for $G_{E_p}$ this model 
is in poor ageement with the SLAC data over the entire $Q^2$ range}  
\cite{andivahis}. Also Holzwarth predicted a linear drop in the 
proton form factor ratio \cite{soliton} before the polarization 
transfer data from Jefferson Lab became available in 2000 \cite{jones,gayou}. 
Since then several theoretical calculations have reproduced this behavior 
\cite{review}. 

\begin{figure}
\centering
\epsfig{file=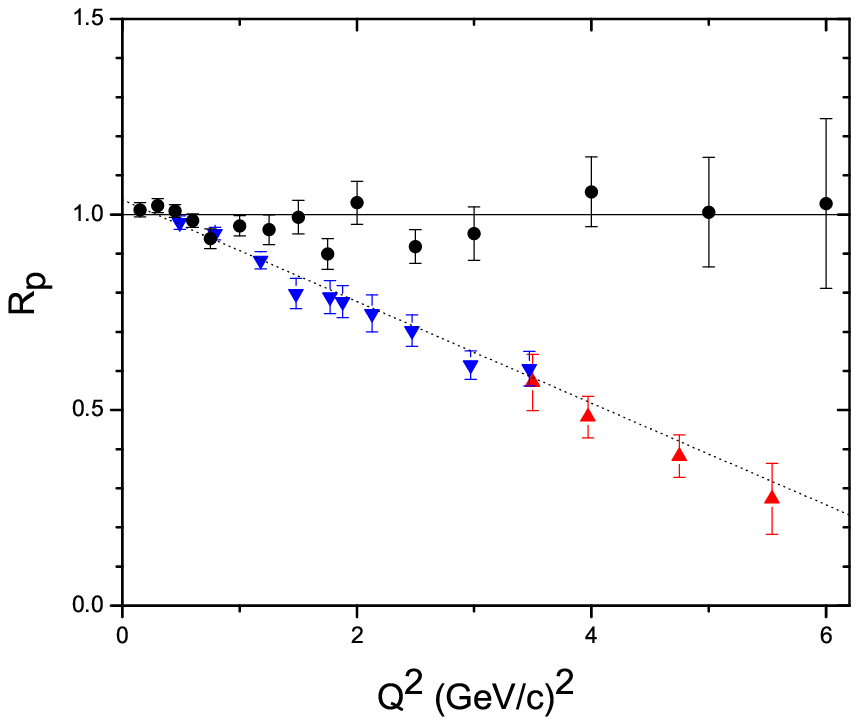}
\caption[]{Comparison of the proton form factor ratio 
$R_p=\mu _{p}G_{E_{p}}/G_{M_{p}}$ obtained from the Rosenbluth separation 
method \protect\cite{walker} (filled circles) and from the polarization 
transfer method \protect\cite{jones} (filled inverted triangles) and 
\protect\cite{gayou} (filled triangles). The dotted and solid lines are 
calculated with Eqs.~(\ref{rp2}) and (\ref{rp1}), respectively.}
\label{gepgmp}
\end{figure}

To illustrate the difference between the two methods, I show in 
Fig.~\ref{gepgmp} the proton form factor ratio $R_p$ obtained from the 
Rosenbluth separation technique \cite{walker} and from the polarization 
transfer experiments \cite{jones,gayou}. In view of the observed discrepancy 
the old Rosenbluth data have been reanalyzed \cite{arrington} and were found 
to be in agreement with form factor scaling of Eq.~(\ref{rp1}).  
In addition, new Rosenbluth experiments have been carried out recently at 
JLab \cite{Christy,Qattan} to try to settle the contradictory experimental 
results. Also in this case, the new results agree within error bars with 
those obtained earlier at SLAC \cite{walker}. Theoretically, radiative 
corrections due to two-photon exchange processes are being investigated 
as a possible source of the observed differences \cite{twophoton,tomasi}. 
The importance of two-photon exchange processes can be tested experimentally 
by measuring the ratio of elastic electron and positron scattering off a 
proton \cite{vepp}.

\section{Vector meson dominance}

The first models to describe the electromagnetic form factors 
of the nucleon are based on vector meson dominance in which it is 
assumed that the photon couples to the nucleon through intermediary 
vector mesons with the 
same quantum numbers as the photon \cite{IJL,hoehler,gari,lomon}. 
In order to take into account the coupling to the vector mesons  
the proton and neutron electromagnetic form factors are expressed 
in terms of the isoscalar, $F^{S}$, and isovector, $F^{V}$, form 
factors as 
\ba
G_{M_{p}} &=&\left( F_{1}^{S}+F_{1}^{V}\right) +\left(
F_{2}^{S}+F_{2}^{V}\right) ~,  \notag \\
G_{E_{p}} &=&\left( F_{1}^{S}+F_{1}^{V}\right) -\tau \left(
F_{2}^{S}+F_{2}^{V}\right) ~,  \notag \\
G_{M_{n}} &=&\left( F_{1}^{S}-F_{1}^{V}\right) +\left(
F_{2}^{S}-F_{2}^{V}\right) ~,  \notag \\
G_{E_{n}} &=&\left( F_{1}^{S}-F_{1}^{V}\right) -\tau \left(
F_{2}^{S}-F_{2}^{V}\right) ~.  \label{ff1}
\ea
In the VMD calculation of 1973 \cite{IJL}, the Dirac form factor was 
attributed to both the intrinsic structure and the meson cloud, and the 
Pauli form factor entirely to the meson cloud. Since this model was 
formulated previous to the development of QCD, no explicit reference 
was made to the nature of the intrinsic structure. In this contribution, 
the intrinsic structure is identified with a three valence quark structure. 
In particular, the question of whether or not there is a coupling to
the intrinsic structure also in the Pauli form factor $F_{2}$ is studied. 
Relativistic constituent quark models in the light-front approach 
\cite{frank,salme} point to the occurrence of such a coupling.
Dimensional counting rules \cite{count} and the development of perturbative 
QCD (p-QCD) \cite{pqcd} has put some
constraints to the asymptotic behavior of the form factors, namely that the
non-spin-flip form factor $F_{1} \rightarrow 1/Q^{4}$ and the spin-flip form
factor $F_{2} \rightarrow 1/Q^{6}$. This behavior has been very recently
confirmed in a perturbative QCD re-analysis \cite{belitsky,brodsky}. The 1973
parametrization, even if it was introduced before the development of p-QCD,
had this behavior. In modifying it, we insist on maintaining the asymptotic
behavior of p-QCD and introduce in $F_{2}^{V}$ a term of the type $%
g(Q^{2})/(1+\gamma Q^{2})$. These considerations lead to the following form of 
the isoscalar and isovector Dirac and Pauli form factors \cite{BI}
\ba
F_{1}^{S}(Q^{2}) &=&\frac{1}{2}g(Q^{2})\left[ 1-\beta _{\omega }-\beta
_{\varphi }+\beta _{\omega }\frac{m_{\omega }^{2}}{m_{\omega }^{2}+Q^{2}}%
+\beta _{\varphi }\frac{m_{\varphi }^{2}}{m_{\varphi }^{2}+Q^{2}}\right] ~, 
\notag \\
F_{1}^{V}(Q^{2}) &=&\frac{1}{2}g(Q^{2})\left[ 1-\beta _{\rho }+\beta _{\rho }%
\frac{m_{\rho }^{2}}{m_{\rho }^{2}+Q^{2}}\right] ~,  \notag \\
F_{2}^{S}(Q^{2}) &=&\frac{1}{2}g(Q^{2})\left[ \left( \mu _{p}+\mu
_{n}-1-\alpha _{\varphi }\right) \frac{m_{\omega }^{2}}{m_{\omega }^{2}+Q^{2}%
}+\alpha _{\varphi }\frac{m_{\varphi }^{2}}{m_{\varphi }^{2}+Q^{2}}\right] ~,
\notag \\
F_{2}^{V}(Q^{2}) &=&\frac{1}{2}g(Q^{2})\left[ \frac{(\mu _{p}-\mu
_{n}-1-\alpha _{\rho })}{1+\gamma Q^{2}}+\alpha _{\rho }\frac{m_{\rho }^{2}}{%
m_{\rho }^{2}+Q^{2}}\right] ~,  \label{ff2}
\ea
with $\mu _{p}=2.793$ $\mu_N$ and $\mu _{n}=-1.913$ $\mu_N$. This 
parametrization insures that the three-quark contribution to the anomalous 
moment is purely isovector, as given by $SU(6)$. For the intrinsic form 
factor a dipole form $g(Q^{2})=(1+\gamma Q^{2})^{-2}$ is used which is 
consistent with p-QCD and in addition coincides with the form used in 
an algebraic treatment of the intrinsic three quark structure \cite{bijker}. 
The values of the meson masses are the standard ones: $m_{\rho}=0.776$ GeV, 
$m_{\omega}=0.783$ GeV, $m_{\varphi}=1.019$ GeV. 

The large width of the $\rho$ meson is crucial for the small $Q^{2}$ behavior 
of the form factors and is taken is taken into account in the same way as in 
\cite{IJL} by the replacement \cite{frazer}
\begin{equation}
\frac{m_{\rho }^{2}}{m_{\rho }^{2}+Q^{2}}\rightarrow \frac{m_{\rho
}^{2}+8\Gamma _{\rho }m_{\pi }/\pi }{m_{\rho }^{2}+Q^{2}+\left( 4m_{\pi
}^{2}+Q^{2}\right) \Gamma _{\rho }\alpha (Q^{2})/m_{\pi }}~,  \label{ff3}
\end{equation}
with 
\begin{equation}
\alpha \left( Q^{2}\right) =\frac{2}{\pi} \sqrt{\frac{4m_{\pi}^{2}+Q^2}{Q^2}} 
\, \ln \left( \frac{\sqrt{4m_{\pi }^{2}+Q^{2}}+\sqrt{Q^{2}}}{%
2m_{\pi }}\right) ~.  \label{ff4}
\end{equation}
For the effective width the same value is taken as in \cite{IJL,wan}:    
$\Gamma _{\rho }=0.112$ GeV. 
For small values of $Q^2$ the form factors are dominated by the meson 
dynamics, whereas for large values the modification from dimensional 
counting laws from perturbative QCD can be taken into account by scaling 
$Q^2$ with the strong coupling constant \cite{gari}
\begin{equation}
Q^{2}\rightarrow Q^{2}\frac{\alpha_s(0)}{\alpha_s(Q^2)} 
\;=\; Q^{2} \frac{\ln \left[ \left( \Lambda ^{2}+Q^{2}\right)
/\Lambda _{\rm QCD}^{2}\right] }
{\ln \left[ \Lambda ^{2}/\Lambda _{\rm QCD}^{2} \right] }~.
\end{equation}
Since this modification is not very important for range 
of $Q^2$ values $0<Q^{2}<10$ GeV$^{2}$ considered in the present 
contribution, it will be neglected in the remainder. 

\begin{figure}
\centering
\epsfig{file=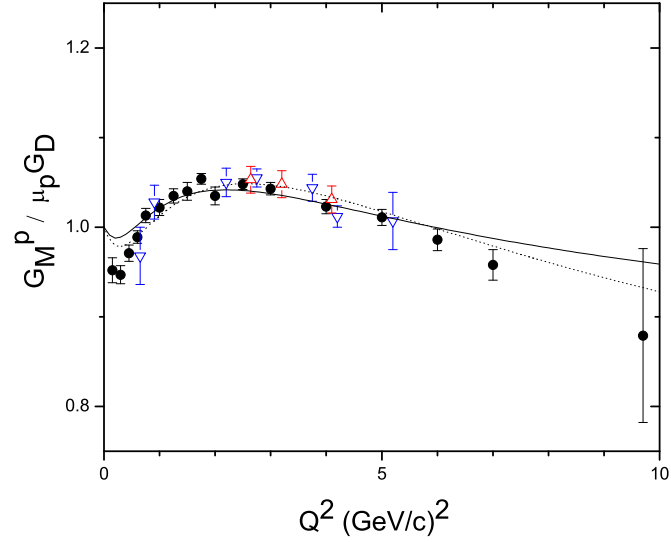}
\caption[]{Comparison between the experimental and theoretical proton 
magnetic form factor $G_{M_p}/\mu_p G_D$ in the space-like region. 
The experimental data included in the fit are taken from 
\protect\cite{walker} (filled circles). Additional data, not included 
in the fit, are taken from \protect\cite{Christy} (open inverted triangles) 
and from \cite{Qattan} (open triangles). The solid line is from the 
present calculation and the dotted line from \protect\cite{IJL}.}
\label{gmp}
\end{figure}

\begin{figure}
\centering
\epsfig{file=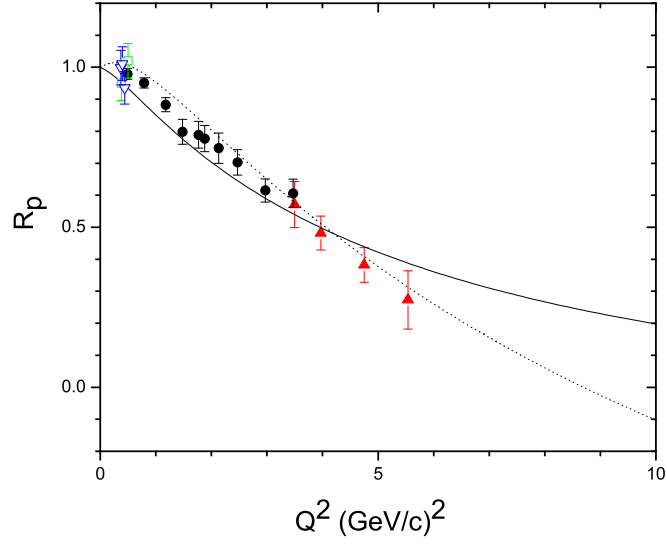}
\caption[]{As Fig.~\ref{gmp}, but for the proton 
form factor ratio $R_p=\mu _{p}G_{E_{p}}/G_{M_{p}}$. 
The experimental data included in the fit are taken from \protect\cite{jones} 
(filled circles) and \protect\cite{gayou} (filled triangles)
Additional data, not included in the fit, are taken from 
\protect\cite{milbrath} (open squares) and \protect\cite{pospi} 
(open inverted triangles).}
\label{rp}
\end{figure}

\begin{figure}
\centering
\epsfig{file=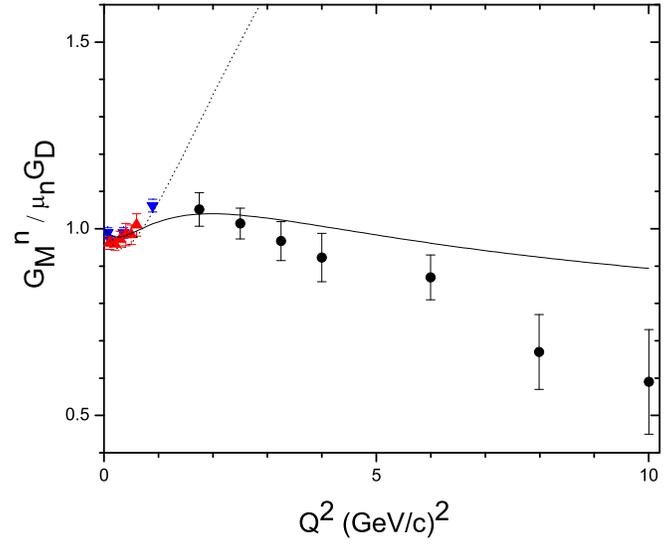}
\caption[]{As Fig.~\ref{gmp}, but for the neutron  
magnetic form factor $G_{M_n}/\mu_n G_D$.  
The experimental data included in the fit are taken from 
\protect\cite{gmnfit}: Rock \& Lung (filled circles) and 
Xu (filled triangles) and \protect\cite{Kubon} (filled inverted triangles).}
\label{gmn}
\end{figure}

\begin{figure}
\centering
\epsfig{file=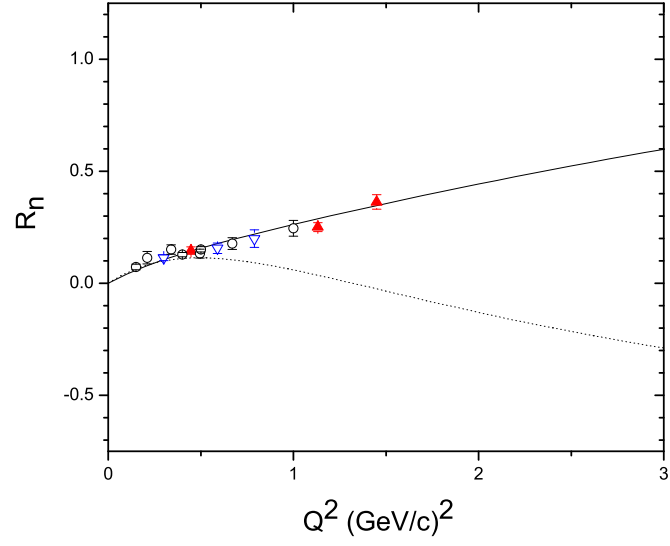}
\caption[]{As Fig.~\ref{gmp}, but for the neutron 
form factor ratio $R_n=\mu _{n}G_{E_{n}}/G_{M_{n}}$. 
The experimental data included in the fit are taken from \protect\cite{madey} 
(filled triangles).  Additional data, not included in the fit, 
are taken from \protect\cite{genfit} 
(open circles) and \cite{glazier} (open inverted triangles).}
\label{rn}
\end{figure}

\begin{figure}
\centering
\epsfig{file=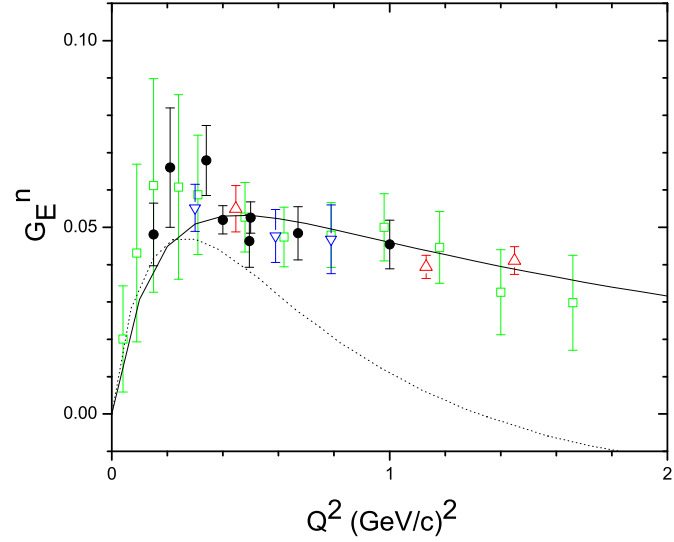}
\caption[]{As Fig.~\ref{gmp}, but for the neutron 
electric form factor $G_{E_n}$. The experimental 
data included in the fit are taken from \protect\cite{genfit} 
(filled circles). Additional data, not included in the fit, are taken from 
\protect\cite{madey} (open triangles) \cite{glazier} (open inverted triangles) 
and \protect\cite{schiavilla} (open squares).}
\label{gen}
\end{figure}

The five remaining coefficients, $\beta_{\rho}$, $\beta_{\omega}$, 
$\beta_{\varphi}$, $\alpha_{\rho}$, $\alpha_{\varphi}$ and the parameter 
$\gamma $ are fitted to recent data on electromagnetic form factors. 
Because of the inconsistencies between different data sets, most notably
between those obtained from recoil polarization and Rosenbluth separation,
the choice of the data plays an important role in the final
outcome. In the present calculation the recoil polarization JLab 
data for the form factor ratios $R_{p}=\mu_{p}G_{E_{p}}/G_{M_{p}}$ and 
$R_{n}=\mu_{n}G_{E_{n}}/G_{M_{n}}$ are used in combination with Rosenbluth 
separation data, mostly from SLAC, for $G_{M_{p}}/\mu_p G_D$ and 
$G_{M_{n}}/\mu_n G_D$, as well as some recent measurements of $G_{E_n}$. 
The data actually used in the fit are quoted in the figure captions to 
Figs.~\ref{gmp}-\ref{gen} and are indicated by filled symbols. 
The values of the fitted parameters are: $\beta_{\rho}=0.512$, 
$\beta_{\omega}=1.129$, $\beta_{\varphi}=-0.263$, $\alpha_{\rho}=2.675$, 
$\alpha_{\varphi}=-0.200$ and $\gamma=0.515$ (GeV/c)$^{-2}$ \cite{BI}. 
These values differ
somewhat from those obtained in the 1973 fit, although they retain most of
their properties, namely a large coupling to the $\omega$ meson in $F_{1}$
and a very large coupling to the $\rho$ meson in $F_{2}$. The spatial
extent of the intrinsic structure is somewhat larger than in \cite{IJL}, 
$\langle r^{2}\rangle ^{1/2}\simeq 0.49$ fm compared to $\simeq 0.34$ fm.

Figs.~\ref{gmp} and~\ref{rp} show a comparison between the calculation 
with the parameters given above and the experimental data for the proton 
magnetic form factor $G_{M_{p}}/\mu _{p}G_{D}$ and the proton form factor 
ratio $R_{p}$. The results of the 1973 calculation, with no direct
coupling to $F_{2}^{V}$, are also shown. One can see that the inclusion of
the direct coupling pushes the zero in $R_{p}$ to larger values of $Q^{2}$
(in \cite{wan} the zero is at $\simeq$ 8 (GeV/c)$^{2}$). Note that any 
model parametrized in terms of $F_{1}$ and $F_{2}$ will produce results for 
$R_{p}$ that are in qualitative agreement with the data, such as a soliton
model \cite{soliton} or relativistic constituent quark models 
\cite{frank,simula}. Perturbation expansions of relativistic effects also 
produce results that go in the right direction \cite{genova}. 

\begin{figure}
\centering
\epsfig{file=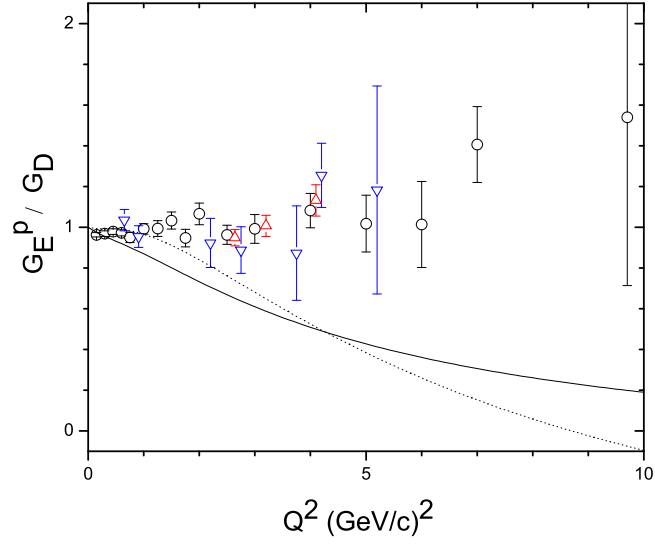}
\caption[]{As Fig.~\ref{gmp}, but for the proton electric form factor 
$G_{E_p}/G_D$. The experimental data, not included in the fit, are obtained 
from the Rosenbluth separation method \protect\cite{walker} (open circles), 
\protect\cite{Christy} (open inverted triangles) and \cite{Qattan} (open triangles).}
\label{gep}
\end{figure}

Figs.~\ref{gmn} and~\ref{rn} show the same comparison, but for the 
neutron. Contrary to the case of the 1973 parametrization, the present
parametrization is in excellent agreement with the neutron data. This is
emphasized in Fig.~\ref{gen} where the electric form factor of the 
neutron is shown and compared with additional data not included in the 
fit. However, as one can see from Figs.~\ref{gmp} and~\ref{rp}, the excellent 
agreement with the neutron data is at the expense of a slight disagreement 
with proton data. Finally, in Fig.~\ref{gep} the results for the proton 
electric form factor is shown in comparison with the experimental data obtained 
from the Rosenbluth technique. To settle the question of consistency between 
proton and neutron space-like data, it is very important to extend the 
existing experimental information to higher values of $Q^2$, more specifically: 
\begin{itemize}
\item Measure the proton form factor ratio $R_p$ beyond 6 (GeV/c)$^{2}$ among 
other things to see whether it indeed goes through zero or not \cite{perdrisat}. 
\item Measure the neutron magnetic form factor $G_{M_{n}}$ beyond 
2 (GeV/c)$^{2}$. This experiment has been carried out at JLab 
and is being analyzed at the moment \cite{brooks}. 
\item Measure the neutron electric form factor $G_{E_{n}}$ beyond 
1.4 (GeV/c)$^{2}$. These experiments will be carried out at JLab \cite{Madey} 
and at MIT-Bates \cite{Gao}. 
\end{itemize}

\subsection{Charge and magnetization radii}

The low $Q^2$ behavior of the electromagnetic form factors provides 
information about the charge and magnetization radii which are related 
to the slope of the electric and magnetic form factors in the origin by               
\ba
\left< r^2 \right>_E &=& -6 \, \left. \frac{dG_E(Q^2)}{dQ^2} \right|_{Q^2=0} ~,
\nonumber\\
\left< r^2 \right>_M &=& -6 \, \frac{1}{\mu} \, \left. \frac{dG_M(Q^2)}{dQ^2} 
\right|_{Q^2=0} ~. 
\ea
In Table~\ref{radii} the radii from the present calculation are compared to 
the experimental values. The proton charge radius and the magnetic radii of 
the proton and the neutron are equal within the error bars. The calculated 
values show the same behavior. However, the absolute values are underpredicted 
by a few percent (note that the radii were not included in the fit). 

\begin{table}
\centering
\caption[]{\small Charge and magnetization radii of the nucleon}
\label{radii}
\vspace{15pt}
\begin{tabular}{lrcc}
\hline
& & & \\
Radius & Present & Experiment & Reference \\
& & & \\
\hline
& & & \\
$\left< r^2 \right>^{1/2}_{E_p}$ &  $0.838$ & $ 0.895 \pm 0.018$ fm     & \cite{Sick} \\
                                 &          & $ 0.890 \pm 0.014$ fm     & \cite{Udem} \\
                                 &          & $ 0.862 \pm 0.012$ fm     & \cite{Simon} \\
& & & \\
$\left< r^2 \right>^{1/2}_{M_p}$ &  $0.825$ & $ 0.855 \pm 0.035$ fm     & \cite{review} \\
& & & \\
$\left< r^2 \right>_{E_n}$       & $-0.100$ & $-0.115 \pm 0.003$ fm$^2$ & \cite{Kopecky} \\
& & & \\
$\left< r^2 \right>^{1/2}_{M_n}$ &  $0.834$ & $ 0.873 \pm 0.011$ fm     & \cite{Kubon} \\
& & & \\
\hline
\end{tabular}
\end{table}

\subsection{Scaling laws}

Dimensional counting rules \cite{count} and results from perturbative QCD 
\cite{pqcd} show that to leading order the ratio of Pauli and Dirac form 
factors scales as $F_2/F_1 \propto 1/Q^2$. However, the values extracted 
from the recently obtained data from polarization transfer experiments 
show that, at least up to $Q^2=5.6$ (GeV/c)$^2$, this ratio seems to scale 
rather as $1/Q$. In the VMD approach of \cite{IJL} this $1/Q$ behavior was 
shown to be transient, {\it i.e.} it is only valid in an intermediate $Q^2$ 
region, but does not hold in the large $Q^2$ limit \cite{iac1}, whereas in 
a p-QCD model with nonzero quark orbital angular momenta wave it was argued 
that the $1/Q$ scaling may be valid to arbitrary large values of $Q^2$ 
\cite{ralston}. In the latter, it was attributed to the presence of the 
nonzero orbital angular momentum components in the 
proton wave function arising from the hadron helicity nonconservation. 
A similar result was obtained in a relativistic constituent quark model 
in which the condition of Poincar\'e invariance induces a violation of 
hadron helicity conservation \cite{miller}. On the other hand, 
in \cite{brodsky} it was argued that the nonzero orbital angular momentum 
contributes to both $F_1$ and $F_2$ in such a way that $F_2/F_1 \propto 1/Q$ 
for intermediate values of $Q^2$ and $\propto 1/Q^2$ for large values of 
the momentum transfer.
 
\begin{figure}
\centering
\epsfig{file=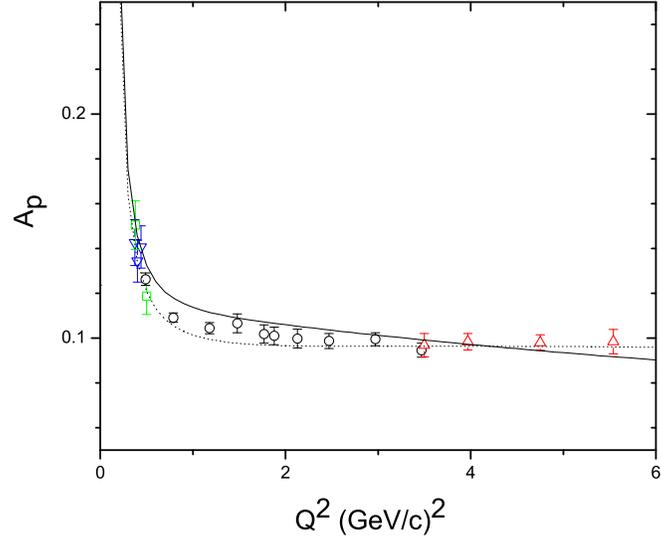}
\caption[]{Comparison between the experimental and theoretical values of 
the ratio $A_p=(Q^2 F_{2,p}/F_{1,p})/\ln^2(Q^2/\Lambda^2_{\rm QCD})$ in 
the space-like region with $\Lambda_{\rm QCD}=0.300$ GeV/c. 
The experimental data are taken from \protect\cite{jones} (open circles), 
\protect\cite{gayou} (open triangles), \protect\cite{milbrath} (open squares) 
and \protect\cite{pospi} (open inverted triangles). The solid line is 
from the present calculation and the dotted line from \protect\cite{IJL}.}
\label{ap}
\end{figure}

\begin{figure}
\centering
\epsfig{file=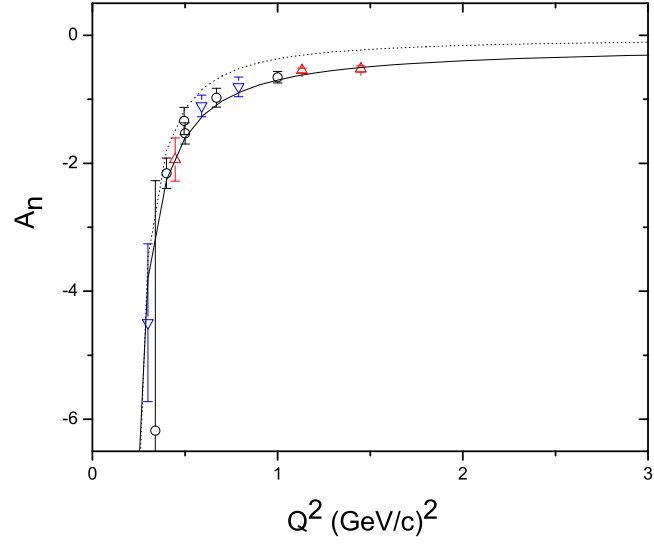}
\caption[]{As Fig.~\protect\ref{ap}, but for the neutron. 
The experimental data are taken from \protect\cite{genfit} 
(open circles), \protect\cite{madey} (open triangles) and 
\cite{glazier} (open inverted triangles).}
\label{an}
\end{figure}

In a recent p-QCD analysis of the Pauli form factor 
the asymptotic behavior of the ratio $F_2/F_1$ was predicted to be 
\cite{belitsky}
\ba
\frac{F_2}{F_1} \propto \frac{\ln^2 (Q^2/\Lambda^2_{\rm QCD})}{Q^2} ~,
\ea
indicating that the quantity 
\ba
A \;=\; \frac{Q^2}{\ln^2 (Q^2/\Lambda^2_{\rm QCD})} \frac{F_2}{F_1} 
\;\propto\; 1 ~,
\ea
does not depend on $Q^2$ in the asymptotic region. The coefficient 
$\Lambda_{\rm QCD}$ is a soft scale related to the size of the nucleon 
$\Lambda_{\rm QCD}=0.300$ GeV/c \cite{review,gari}.  
A similar form was obtained in a study of light-front wave functions 
\cite{brodsky}. 
Figs.~\ref{ap} and~\ref{an} show this ratio for the proton and neutron, 
respectively. The data seem to approach a constant value, even though 
the domain of validity of p-QCD is expected to set in at much higher 
values of $Q^2$. In order to understand the onset of the perturbative 
region of QCD, it is crucial to extend the polarization transfer 
measurements to larger values of $Q^2$ both for the proton and the 
neutron. 

\section{Time-like form factors}

Time-like form factors are important for a global understanding 
of the structure of the nucleon \cite{hammer,dub,egle,BCHH,TG}. 
In the space-like region ($Q^2 > 0$) the electromagnetic form factors 
can be studied through electron scattering, whereas in the time-like 
region ($Q^2 < 0$) they can be measured through the creation or 
annihilation of a nucleon-antinucleon pair. 
The time-like structure of the nucleon 
form factors has recently been analyzed \cite{wan} in the VMD model 
of \cite{IJL}. Here the same approach is used to analyze the time-like 
structure of the form factors. The method consists in analytically 
continuing the intrinsic structure to \cite{BI} 
\begin{equation}
g(q^{2})=\frac{1}{(1-\gamma e^{i\theta }q^{2})^{2}}~,  \label{ff5}
\end{equation}
where $q^{2}=-Q^{2}$ and $\theta $ is a phase. The contribution of the 
$\rho$ meson is analytically continued for $q^{2}>4m_{\pi }^{2}$ as 
\cite{frazer}
\begin{equation}
\frac{m_{\rho }^{2}}{m_{\rho }^{2}-q^{2}}\rightarrow \frac{m_{\rho
}^{2}+8\Gamma _{\rho }m_{\pi }/\pi }{m_{\rho }^{2}-q^{2}+(4m_{\pi
}^{2}-q^{2})\Gamma _{\rho }[\alpha (q^{2})-i\beta (q^{2})]/m_{\pi }}~,
\label{ff6}
\end{equation}
where 
\begin{eqnarray}
\alpha (q^{2}) &=&\frac{2}{\pi }\sqrt{\frac{q^{2}-4m_{\pi}^{2}}{q^{2}}}
\, \ln \left( \frac{\sqrt{q^{2}-4m_{\pi }^{2}}+\sqrt{q^{2}}}
{2m_{\pi }}\right) ~,  \notag \\
\beta (q^{2}) &=&\sqrt{\frac{q^{2}-4m_{\pi}^{2}}{q^{2}}}~.  
\label{ff7}
\end{eqnarray}

\begin{figure}
\centering
\epsfig{file=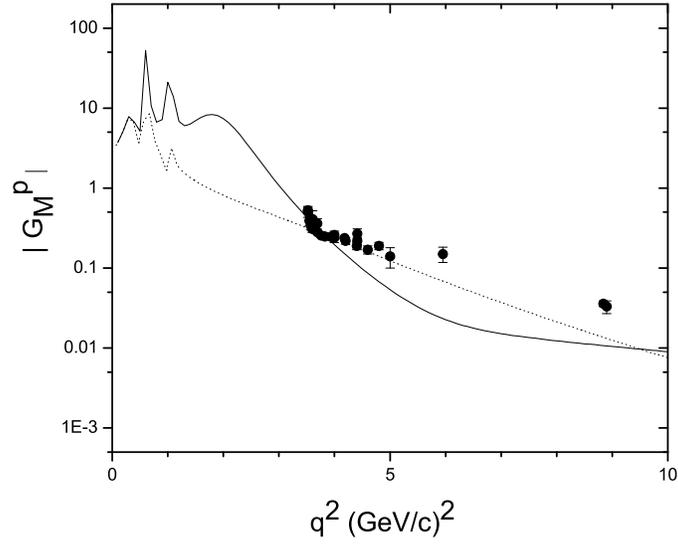}
\caption[]{Comparison between experimental and theoretical proton magnetic 
form factor $|G_{M_{p}}|$ in the time-like region. 
The experimental values are taken from \protect\cite{ptime} 
under the assumption $|G_{E_{p}}|=|G_{M_{p}}|$. 
The solid lines are from the present analysis and the dotted lines 
from \protect\cite{wan}.}
\label{gmptime}
\end{figure}

\begin{figure}
\centering
\epsfig{file=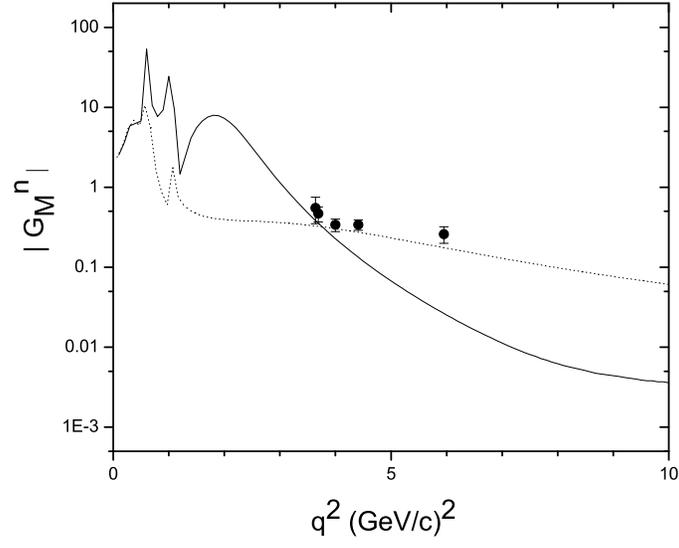}
\caption[]{As Fig.~\ref{gmptime}, but for the neutron magnetic form factor 
$|G_{M_{n}}|$. The experimental values are taken from \protect\cite{ntime} 
under the assumption $|G_{E_{n}}|=0$.}
\label{gmntime}
\end{figure}

\begin{figure}
\centering
\epsfig{file=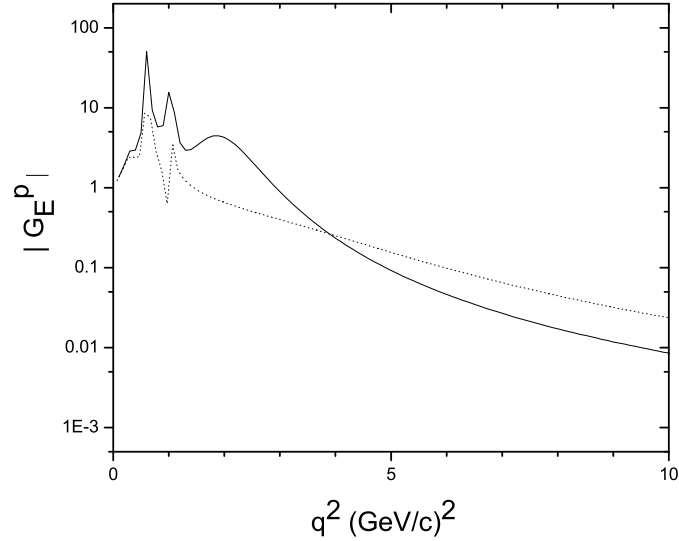}
\caption[]{As Fig.~\ref{gmptime}, but for the proton electric form factor 
$|G_{E_{p}}|$.}
\label{geptime}
\end{figure}

\begin{figure}
\centering
\epsfig{file=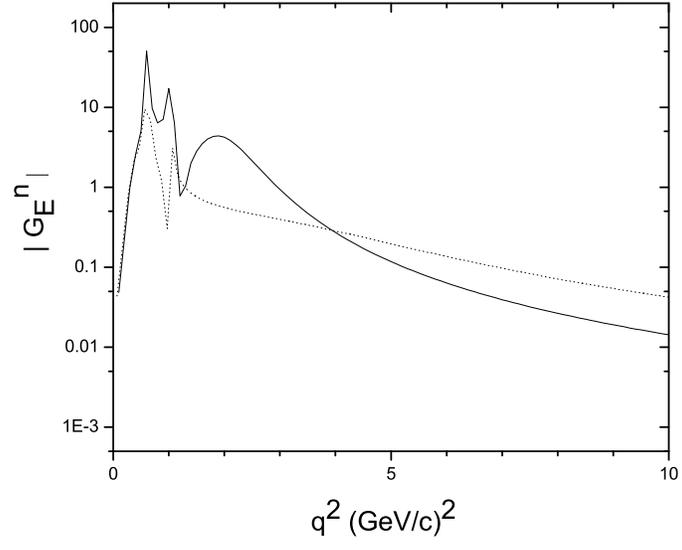}
\caption{As Fig.~\ref{gmptime}, but for the neutron electric form factor 
$|G_{E_{n}}|$.}
\label{gentime}
\end{figure}

The results for the time-like form factors are shown in Figs.~\ref{gmptime} 
and~\ref{gmntime}. The phase $\theta$ is 
obtained from a best fit to the proton data: $\theta =0.397$ rad 
$\simeq 22.7^{\circ}$, compared to the value $\simeq 53^{\circ}$
obtained in \cite{wan}. It should be noted that the correction to the large 
$q^{2}$ data discussed in \cite{wan} has not been done in these figures. One
can see from these figures that while the proton form factor, $|G_{M_{p}}|$, 
obtained from analytic continuation of the present parametrization is in 
marginal agreement with data, the neutron form factor, $|G_{M_{n}}|$, is in 
major disagreement. This result points once more to the inconsistency 
between neutron space-like and time-like data already noted in 
\cite{wan,hammer}. A remeasurement of the neutron 
time-like data could help to resolve this inconsistency.  
The result presented here is in contrast with the analysis of \cite{wan} 
that was in good agreement with both proton and neutron
time-like form factors. In Figs.~\ref{geptime} and~\ref{gentime}, the electric
form factors $|G_{E_{p}}|$ and $|G_{E_{n}}|$ are shown for future use
in the extraction of $|G_{M_{p}}|$ and $|G_{M_{n}}|$ from the data. These 
figures show that the assumptions $|G_{E_{p}}|=|G_{M_{p}}|$ and 
$|G_{E_{n}}|=0$ used in the extraction of the magnetic form factors from the
experimental data, are not always justified. 

\section{Summary and conclusions}

In this contribution, a simultaneous study of the space- and time-like 
data on the electromagnetic form factors of the nucleon was presented. 
The analysis was carried out in a two-component model, which consists of 
an intrinsic (three-quark) structure and a meson cloud whose effects were 
taken into account via VMD couplings to the $\rho$, $\omega$ and $\varphi$ 
mesons. The difference with an earlier calculation lies in the treatment 
of the isovector Pauli form factor $F_2^V$. The inclusion of an intrinsic 
component in $F_2^V$ shows a considerable improvement for the space-like 
neutron form factors. Since the form factors are related by isospin symmetry, 
the excellent description description of the neutron form factors is at the 
expense of a slight disagreement for the proton space-like data. 
The picture emerging from the present study is that of an intrinsic 
structure slightly larger in spatial extent than that of \cite{wan}, 
$\langle r^{2}\rangle ^{1/2}\simeq 0.49$ fm instead of $0.34$
fm, and a contribution of the meson cloud ($q\bar{q}$ pairs) slightly
smaller in strength than that of \cite{wan}, $\alpha _{\rho }=2.675$
instead of $3.706$. 

The present calculation is not able to describe the neutron time-like data in a 
satisfactory way. In effect, a simultaneous description of both the space- 
and time-like data for the neutron has encountered serious difficulty in 
the literature \cite{hammer,wan,BI}. It is of the utmost importance to 
extend the experimental information on the neutron form factors to be able 
to resolve the observed discrepancies and to obtain a consistent description 
of all electromagnetic form factors of the nucleon in both the 
space- and time-like regions. 

Recently it was shown \cite{BCHH,TG} that 
the angular dependence of the single-spin and double-spin polarization 
observables provides a sensitive test of model of nucleon form factors 
in the time-like region. These polarization observables allow to 
distinguish between different models of nucleon electromagnetic form 
factors, even though they fit equally well the nucleon form factors 
in the space-like region. More work in this direction is in progress. 

\section*{Acknowledgments}

This work was supported in part by Conacyt, Mexico. It is a pleasure to thank 
Franco Iachello and Kees de Jager for stimulating discussions.

\clearpage

\end{document}